\documentclass[preprint,floats,aps,epsfig,nofootinbib,amssymb]{revtex4}
\usepackage{axodraw}
\usepackage{slashed}
\usepackage{graphicx}
\usepackage{epsfig}
\usepackage{subfigure}
\usepackage{dcolumn}
\usepackage{bm}
\def\gsim{\lower0.5ex\hbox{$\:\buildrel >\over\sim\:$}}
\def\lsim{\lower0.5ex\hbox{$\:\buildrel <\over\sim\:$}}

\begin{document}


\title{Neutrino Masses and a TeV Scale Seesaw Mechanism}

\author{\bf Wei Chao}
\email{chaow@pku.edu.cn}

\affiliation{Center for High Energy Physics, Peking university,
Beijing 100871, China \vspace{2.5cm} }

\begin{abstract}
A simple extension of the Standard Model providing TeV scale seesaw
mechanism is presented. Beside the Standard Model particles and
right-handed Majorana neutrinos, the model contains a singly charged
scalar, an extra Higgs doublet and three vector like singly charged
fermions. In our model, Dirac neutrino mass matrix raises only at
the loop level. Small but non-zero Majorana neutrino masses come
from integrating out heavy Majorana neutrinos, which can be at the
TeV scale. The phenomenologies of the model are investigated,
including scalar mass spectrum, neutrino masses and mixings, lepton
flavor violations, heavy neutrino magnetic moments as well as
possible collider signatures of the model at the LHC.
\end{abstract}
\maketitle

\section{Introduction}

The observation of neutrino oscillations \cite{sno, abcc, kamland,
k2k} has revealed that neutrinos have small but non-zero masses and
lepton flavors are mixed, which can not be accommodated in the
Standard Model (SM) without introducing extra ingredients. As such,
neutrino physics offer an exciting window into new physics beyond
the SM. Perhaps the most attractive approach towards understanding
the origin of small neutrino masses is using the dimension-five
weinberg operator \cite{weinberg}:
\begin{eqnarray}
{1 \over 4 } \kappa_{gf}^{}
\overline{\ell_{Lc}^C}^g\varepsilon_{cd}^{} \phi_d^{} \ell_{Lb}^f
\varepsilon_{ba}^{} \phi_a^{} + {\rm h.c.} \; ,
\end{eqnarray}
which comes from integrating out new  superheavy  particles.

A simple way to obtain the operator in Eq. (1) is through the Type-I
seesaw mechanism \cite{seesawI}, in which three right-handed
neutrinos with large Majorana masses are introduced to the SM. Then
three active neutrinos may acquire tiny Majorana masses through the
Type-I seesaw formula, i.e., the mass matrix of light neutrinos is
given by $M_\nu^{}= - M_D^{} M_R^{-1} M_D^T$, where $M_D^{}$ is the
Dirac mass matrix linking left-handed light neutrinos to
right-handed heavy neutrinos and $M_R^{}$ is the mass matrix of
heavy Majorana neutrinos. Actually, there are three tree-level
seesaw scenarios (namely type-I, Type-II \cite{seesawII} and
Type-III \cite{seesawIII} seesaw mechanisms) and one loop-level
seesaw scenario (namely Ma \cite{ma} model), which may lead to the
effective operator in Eq. (1).

Although seesaw mechanisms can work naturally to generate Majorana
neutrino masses, they lose direct testability on the experimental
side. A direct test of seesaw mechanism would involve the detection
of these heavy seesaw particles at a collider and the measurement of
their Yukawa couplings with the electroweak doublets. In the
canonical seesaw mechanism, heavy seesaw particles turn out to be
too heavy, i.e., $10^{14 \sim 16 }$ ${\rm GeV}$, to be
experimentally accessible. One straightforward way out is to lower
the seesaw scale ``by hand" down to the TeV scale, an energy
frontier to be explored by the Large Hadron Collider (LHC). However
this requires the structural cancellation between the Yukawa
coupling texture and the heavy Majorana mass matrix, i.e. $M_D
M_R^{-1} M_D^T \approx 0$ \cite{Pilaftsis, early, smirnov, Han,
spanish, tev type-II} at the tree level, and is thus unnatural!

To solve this unnaturalness problem, we propose a novel TeV-scale
seesaw mechanism in this paper. The model includes, in addition to
the SM fields and right-handed Majorana neutrinos, a charged scalar
singlet, an extra Higgs doublet and three vector like singly charged
fermions. Due to $Z_2^{}$ discrete flavor symmetry, right-handed
Majorana neutrinos don't couple to left-handed lepton doublets, such
that Dirac mass matrix only raises at the loop level and is
comparable with the charged lepton mass matrix. This drives down the
seesaw scale to  the TeV, and thus the model is detectable at the
LHC.

The paper is organized as follows: In section II, we describe our
model. Section III is devoted to investigate the phenomenologies of
the model, including neutrino masses and mixings, lepton flavor
violations, transition magnetic moments of heavy Majorana neutrinos
as well as possible collider signatures. We conclude in Section IV.
An alternative settings to the model is presented in appendix A.

\section{The model}

In our model, we extend the SM by introducing three right-handed
Majorana neutrinos $N_R^{}$, three singly charged vector-like
fermion $S_L^{}, S_R^{}$, an extra Higgs doublet $H_n^{}$, a singly
charged scalar $\Phi$ as well as discrete $Z_2^{}$ flavor symmetry.
The $Z_2^{}$ charges for these fields is given in table I. Due to $
Z_2^{}$ symmetry, right-handed neutrinos don't couple to SM Higgs.
\begin{table}[htbp]
\caption{ $Z_2^{}$ charges of particles. } \centering
\begin{tabular}{c|c|c|c|c|c|c|c|c}
\hline  fields & $\ell_L^{}$ & $e_R^{}$ & $N_R^{}$ & $S_L^{}$ & $S_R^{}
$ & $H$ & $H_n^{}$ & $\Phi$ \\
\hline $Z_2^{}$ & +1 & +1 & -1 & -1 & +1 & +1 & +1 & +1 \\
\hline
\end{tabular}
\end{table}

As a result the new lagrangian can be written as
\begin{eqnarray}
{\cal L}_{\rm N}^{} = V(H, H_n^{}, \Phi)  -{\overline{\ell_L^{}}}
Y_S^{} H S_R^{} -\overline{S_L^{}} M_S^{} S_R^{} - Y_N^{}
\overline{S_L^{}} \Phi N_R^{}- {1 \over 2 } \overline{N_R^C} M_R^{}
N_R^{} + {\rm h.c.} \; ,
\end{eqnarray}
where $Y_S^{}$ and $Y_N^{}$ are new Yukawa couplings, $M_S^{}$ and
$M_R^{}$ are mass matrices of $S$ and $N_R^{}$, respectively.
$Z_2^{}$ symmetry is explicitly broken by $\overline S_L^{} M_S^{}
S_R^{}$ term. It can be recovered by adding an extra scalar singlet
$\eta$, with $Z_2^{}$ charge $-1$ and Yukawa coupling
$\overline{S_L^{}} \eta S_R^{}$. We will not consider Yukawa
couplings $\overline{\ell_L^{}} H_n^{} e_R^{}$ and
$\overline{\ell_L^{}} H_n^{} S_R^{}$, which can be forbidden by
another $Z_2^{'}$ symmetry. The following is the full Higgs
potential:
\begin{eqnarray}
V&=& - m_1^{2} H^\dagger H^{} - m_2^{2} H_n^\dagger H_n^{} - m_3^2
\Phi^\dagger \Phi + \lambda_1^{} (H^\dagger H)^2 + \lambda_2^{}
(H_n^\dagger H_n^{})^2   \nonumber   \\
&&+ \lambda_3^{} (H^\dagger H)(H_n^\dagger H_n^{}) + { \lambda_4^{}
\over 4 } (H^\dagger H_n^{}+ H_n^\dagger H )^2 -
{\lambda_5^{} \over 4} (H^\dagger H_n^{}- H_n^\dagger H )^2 \nonumber \\
&& + \lambda_6^{} (\Phi^\dagger \Phi ) (H^\dagger H ) + \lambda_7^{}
(\Phi^\dagger \Phi ) ( H_n^\dagger H_n^{} )+ \left[ \lambda_{n}^{}
\Phi (H^T i \sigma_2^{} H_n^{}) + { \rm h.c. } \right] \; .
\end{eqnarray}
We define $\langle H \rangle =v_1^{}/\sqrt{2}$ and $\langle H_n^{}
\rangle = v_2^{}/\sqrt{2}$. After imposing the conditions of global
minimum, one finds that
\begin{eqnarray}
v_1^2 ={ \lambda_2^{} m_1^2- \lambda_o^{} m_2^2 \over \lambda_1^{}
\lambda_2^{} -\lambda_o^2} \; ; \hspace{2cm} v_2^2 = {\lambda_1^{}
m_2^2 -\lambda_o^{} m_1^2 \over \lambda_1^{} \lambda_2^{} -
\lambda_o^2} \; ,
\end{eqnarray}
where  $\lambda_o^{}= 1/2 (\lambda_3^{} + \lambda_4^{})$.

In the basis $(h^-, h_n^-,  S^-)$,  we can derive the mass matrix
for charged scalars:
\begin{eqnarray}
M_{\rm C}^{} =\left( \matrix{-\lambda_4^{} v_2^2 & \lambda_4^{}
v_1^{} v_2^{} & \sqrt{2} \lambda_n^{} v_2^{} \cr \lambda_4^{} v_1^{}
v_2^{} &- \lambda_4^{} v_1^2 & - \sqrt{2} \lambda_n^{} v_1^{} \cr
\sqrt{2} \lambda_n^{} v_2^{} & - \sqrt{2} \lambda_n^{} v_1^{} & \rho
} \right) \; ,
\end{eqnarray}
where $\rho \equiv -2 m_3^2 +  \lambda_6^{} v_1^2 + \lambda_7^{}
v_2^2$. $M_{\rm C}^{}$ can be diagonalized by the $3 \times 3$
unitary transformation $V$: $V^\dagger M_{\rm C}^{} V^* = {\rm diag}
(M_{G^+}^{}, M_{H^+}^{}, M_{S^+}^{} )$. The mass eigenvalues for
these charged scalars are then
\begin{eqnarray}
M_{ G^+}^{}&=&0 \; ; \nonumber \\
M_{ H^+}&=& {1\over 2}\left| B - \sqrt{B^2 + 4(v_1^2+ v_2^2)(\rho
\lambda_4^{} + 2 \lambda_n^2)} \right| \; ; \nonumber \\
M_{ S^+}^{} &=& {1\over 2}\left| B + \sqrt{B^2 + 4(v_1^2+
v_2^2)(\rho \lambda_4^{} + 2 \lambda_n^2)} \right| \; ,
\end{eqnarray}
where $B\equiv \rho-\lambda_4^{}( v_1^2 + v_2^2)$. Here $G^+$ is the
SM goldstone boson. We also derive the mass matrix for CP-even
scalars in the basis $(h, h_n^{})^T$ and CP-odd scalars in the basis
$(G, G_n^{})^T$:
\begin{eqnarray}
M_{\rm N}^{}= \left(  \matrix{2 \lambda_1^{} v_1^2 & \lambda_o^{}
v_1^{} v_2^{} \cr \lambda_o^{}v_1^{} v_2^{} & 2 \lambda_2^{} v_2^2
}\right) \; ; \hspace{1cm} M_{\rm G}^{} = \left(
\matrix{-\lambda_s^{} v_2^2 & \lambda_s^{} v_1^{} v_2^{} \cr
\lambda_s^{} v_1^{} v_2^{} & -\lambda_s^{} v_1^2} \right) \; ,
\end{eqnarray}
where $\lambda_s^{} = 1/2(\lambda_4^{}-\lambda_5^{})$.

We also derive the masses for gauge bosons, which are $M_W^2=  g^2
(v_1^2 + v_2^2)/4$ and $M_Z^2=  g^2 (v_1^2 + v_2^2)/ 4\cos^2
\theta_w^{}$, separately. Such that electroweak precision observable
$ \rho \equiv M_W^2/ M_Z^2 \cos^2 \theta_w =1$ in our theory. Our
scalar field sector is similar to that in Zee model \cite{zee}. We
present in appendix A a different setting to the particle contents,
by replacing scalar singlet with triplet.

\section{Phenomenological analysis}
In this section, we devote to investigate some phenomenological
implications of our model. We focus on (A) neutrino masses and
mixings; (B)lepton flavor violations; (C) electromagnetic properties
and (D) collider signatures of heavy majorana neutrinos, which will
be deployed in the following:

\subsection{Neutrino masses and lepton mixing martrix}

In our model, there is no Dirac neutrino mass term at the tree
level. However we can derive a small Dirac neutrino mass matrix at
the loop level. The relevant feynman diagram is shown in Fig. 1.
\begin{figure}[h]
\includegraphics[scale=0.4]{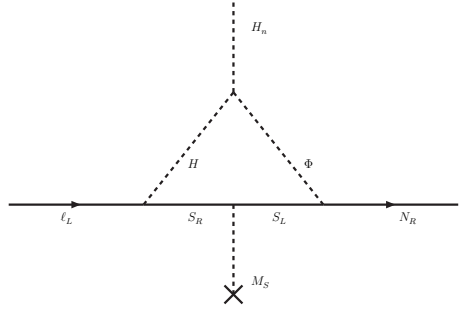}
\caption{One-loop correction to the Dirac neutrino mass matrix.}
\end{figure}

A direct calculation results in
\begin{eqnarray}
\left(M_D^{}\right)_{ab}^{\rm loop} = { \lambda_n^{} \langle H_n^{}
\rangle \over 32 \pi^2} \left(Y_S^{}\right)_{ac}^{}(
Y_N^{\dagger})_{cb}^{} M_{S_c}^{} {\cal F } \left( M_H^2, M_\Phi^2,
M_{S_c}^2 \right) \; ,
\end{eqnarray}
where the one loop function appearing in upper equation is given by
\begin{eqnarray}
{\cal F}(m_1^2, m_2^2, m^2) = {1 \over m_1^2-m_2^2 } \left( {\ln
\beta_1 \over \beta_1 -1 } -{\ln \beta_2 \over \beta_2-1} \right) \;
,
\end{eqnarray}
with $\beta_i = m^2/ m_i^2$. When $m_1^{}= m_2^{}=m$, ${\cal F}$
reduces to $1 / 2m^2 $.

Here $M_D^{\rm loop}$ is the Dirac neutrino mass matrix linking the
left and right hand neutrinos, which only raises at the loop level
in our model. If neutrinos are Dirac particles, then Eq. (8) is just
neutrino mass formula, whose predication must be consistent with
present neutrino oscillation data. In this paper, we assume that
neutrinos are Majorana particles, i.e., left-handed and right-handed
neutrinos have different mass eigenvalues. Then three active
neutrino masses can be generated from seesaw mechanism. In this
case, we can write down the $6 \times 6$ neutrino mass matrix:
\begin{eqnarray}
{\cal M}= \left( \matrix{ 0 & M_D^{\rm loop} \cr M_D^{{\rm loop}T }
& M_R^{}} \right) \; ,
\end{eqnarray}
which can be diagonalized by the unitary transformation ${\cal
U}^\dagger {\cal M} {\cal U}^* = \widehat{ {\cal  M}}$; or
explicitly,
\begin{eqnarray}
\left(  \matrix{V & R \cr S & T}\right)^\dagger  \left( \matrix{ 0 &
M_D^{\rm loop} \cr M_D^{{\rm loop}T } & M_R^{}} \right) \left(
\matrix{V & R \cr S & T}\right)^* = \left(  \matrix{ \hat M_\nu{} &
0 \cr 0 & \hat M_N^{}}\right) \; .
\end{eqnarray}
Given $M_D^{\rm loop} \ll M_R^{}$, the light Majorana neutrino mass
formula is then $M_\nu^{} = - M_D^{\rm loop} M_R^{-1} {M_D^{\rm loop
}}^T$. Notice that Dirac neutrino mass matrix is suppressed by loop
factor, we assume $M_D^{\rm loop} \sim {\cal O} (\rm MeV)$, which
will not cause any fine-tune problem. Then, to generate
electron-volt scale active neutrino masses, heavy Majorana neutrinos
would be of the order of several hundred GeV.

We also obtain the charged lepton mass matrix in the basis $ (E_L^{}
, S_L^{} )^T$,
\begin{eqnarray}
M_{\ell} =\left( \matrix{M_E^{} &  M_C^{} \cr 0  & M_S^{} }
\right)=\left( \matrix{ \textbf{I} & M_C^{} M_S^{-1} \cr 0 &
\textbf{I}}\right) \left( \matrix{M_E^{} & 0 \cr 0 & M_S^{}}\right)
 \; ,
\end{eqnarray}
where $M_E^{} = v / \sqrt{2}Y_E^{} $ and $M_C^{}= v/ \sqrt{2} Y_S^{}
$.

According to Eqs. (11) and (12), we may derive the lepton mixing
matrix (MNS), which comes from the mismatch between the
diagonalizations of the neutrino mass matrix and charged lepton mass
matrix, i.e., $U=V_e^\dagger V_\nu^{}$:
\begin{eqnarray}
U\approx ( \textbf{1} - {1 \over 2} |M_C^{} M_S^{-1}|^2 ) V \; .
\end{eqnarray}
As a result, the effective charged and neutral current interactions
for charged leptons can be written as
\begin{eqnarray}
-{\cal L}_{ \rm CC}^{} &\approx& {g\over \sqrt{2}}
\overline{e_{\alpha}^{}} \gamma^\mu P_L^{} U_{\alpha i}^{}
\nu_i^{} W_\mu^{} + {\rm h.c.} \; ; \\
-{\cal L}_{\rm NC}^{} &\approx& {g \over \cos
\theta_w^{}}\overline{e_\alpha^{}} \gamma^\mu \left[(U^{}
U^\dagger)_{\alpha \beta}^{}(-{1\over 2} + \sin^2 \theta_w) P_L^{} +
\delta_{\alpha \beta}^{} \sin^2 \theta_w^{} P_R^{}\right] e_\beta^{}
Z_\mu^{} \; .
\end{eqnarray}

The MNS matrix in Eq. (13) is non-unitary, which is mainly because
the large mixing between charged leptons and heavy vector like
fermions.  To a better degree of accuracy, we have $UU^\dagger
\approx \textbf{1} - |M_C^{} M_S^{-1}|^2$. A global analysis of
current neutrino oscillation data and precision electroweak data
(e.g., on the invisible width of the $Z^0$ boson, universality tests
and rare decays) has yield quite strong constraints on the unitarity
of $U$. Translating the numerical results of Refs. \cite{zerodis,
extest1, zhizhong, extest2} into the restriction on $|M_C^{}
M_S^{-1}|^2$, we have
\begin{eqnarray}
|M_C^{} M_S^{-1}|^2 = \left( \matrix{< 1.1\cdot 10^{-2} & < 7.0\cdot
10^{-5} & <1.6\cdot 10^{-2} \cr < 7.0\cdot 10^{-5} & < 1.0\cdot
10^{-2} & < 1.0 \cdot 10^{-2} \cr <1.6\cdot 10^{-2} & <1.0\cdot
10^{-2} & < 1.0\cdot10^{-2}} \right) \; ,
\end{eqnarray}
at the $90\%$ confidence level. In addition, interactions in Eqs.
(14) and (15) will lead to tree level lepton flavor violations (as
can be seen in Eq. (15)) and `` zero distance" effects
\cite{zerodis} in neutrino oscillations, which can be verified in
the future long baseline neutrino oscillation experiments.

\subsection{Lepton flavor violations (LFV)}

Notice that the emergence of big unitary violation of MNS matrix can
lead to observable LFV effects. In this subsection, we investigate
constraints on parameter space from LFV processes.

In our model, $\ell_i^{} \rightarrow 3 \ell_j^{}$ may occur at the
tree level, just like the case in type III seesaw mechanism. The
branching ratios for the $\mu \rightarrow 3 e $ can be given by
\begin{eqnarray}
{\rm BR} (\mu \rightarrow  e^- e^+ e^-) = |UU^\dagger|_{ e
\mu}^2\left[ |UU^\dagger|_{ee}^{} (-1+ 2 \sin^2 \theta_w^{})^2
+4\sin^4 \theta_w^{} \right] \Omega \; .
\end{eqnarray}
Here $\Omega$ is the final states phase space integration
\begin{eqnarray}
\Omega = \int_{z_d^{}}^{z_u^{}} \sqrt{1 - {4z_0^{} \over z }} \left[
-2 z^2 + (1+ 3 z_0^{})z -4 z_0^{}(1+ z_0^{}) + {2 z_0^{}(1-z_0^{})^2
\over z} \right] \lambda^{1\over 2} (1, z_0^{}, z) d z \; ,
\end{eqnarray}
where $z_d^{} = 4 z_0^{}$, $z_u^{} = (1-\sqrt{z_0^{}})^2$, $z_0^{} =
m_e^2 / m_\mu^2$ and $\lambda(x, y, z)=x^2 + y^2 + z^2 -2
xy-2xz-2yz$.

Radiative decays, i.e., $\ell_i^{} \rightarrow \ell_j^{} + \gamma$
occur at one-loop level.  The branching ratios for these processes
can be written as
\begin{eqnarray}
{\rm BR}(\ell_\beta^{} \rightarrow \ell_\alpha^{} \gamma ) = {3 \pi
\alpha  \over 16 G_{\rm F}^2 m_\beta^4} \left | \sum_i^{} Y_{S\alpha
i}^{} Y_{S\beta i}^* {\cal S}_i^{} \right |^2 {\rm BR}(\ell_\beta
\rightarrow \nu_\beta^{} \ell_\alpha^{} \bar \nu_\alpha^{}) \; ,
\end{eqnarray}
with
\begin{eqnarray}
{\cal S }_\rho^{} = { \sqrt{z_\alpha^{} z_\beta^{} } (1- z_h^{} ) -2
z_\beta^{}(1-z_h^{} ) + 2 \sqrt{z_\alpha^{} z_\beta^{}} \over
(1-z_h^{})^2} - {z_\beta^{}(1-z_h^{})-\sqrt{z_\alpha z_\beta} \over
(1-z_h^{})^3} \ln z_h^{} \; ,
\end{eqnarray}
where $z_\alpha^{} = {m_\alpha^{2} / M_{Si}^{2}} \; ;$ $z_\beta^{} =
{m_\beta^{2}/ M_{Si}^{2}}  \; ;$ $z_h^{} = {M_h^{2} / M_{Si}^{2}} \;
.$
\begin{figure}[t]
\subfigure[]{\epsfig{file=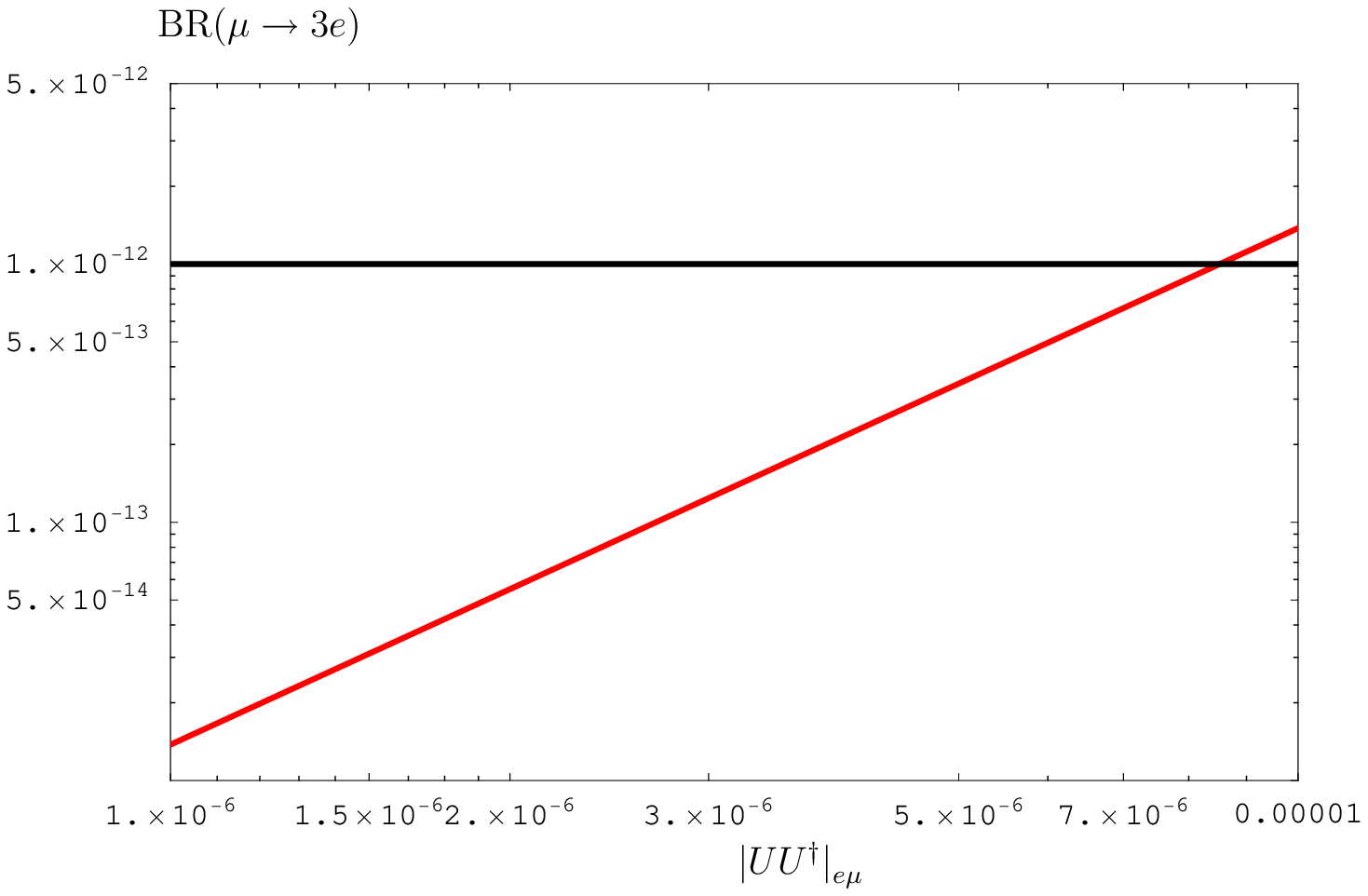,height=5.5cm,width=7.5cm,angle=0}}
\subfigure[]{\epsfig{file=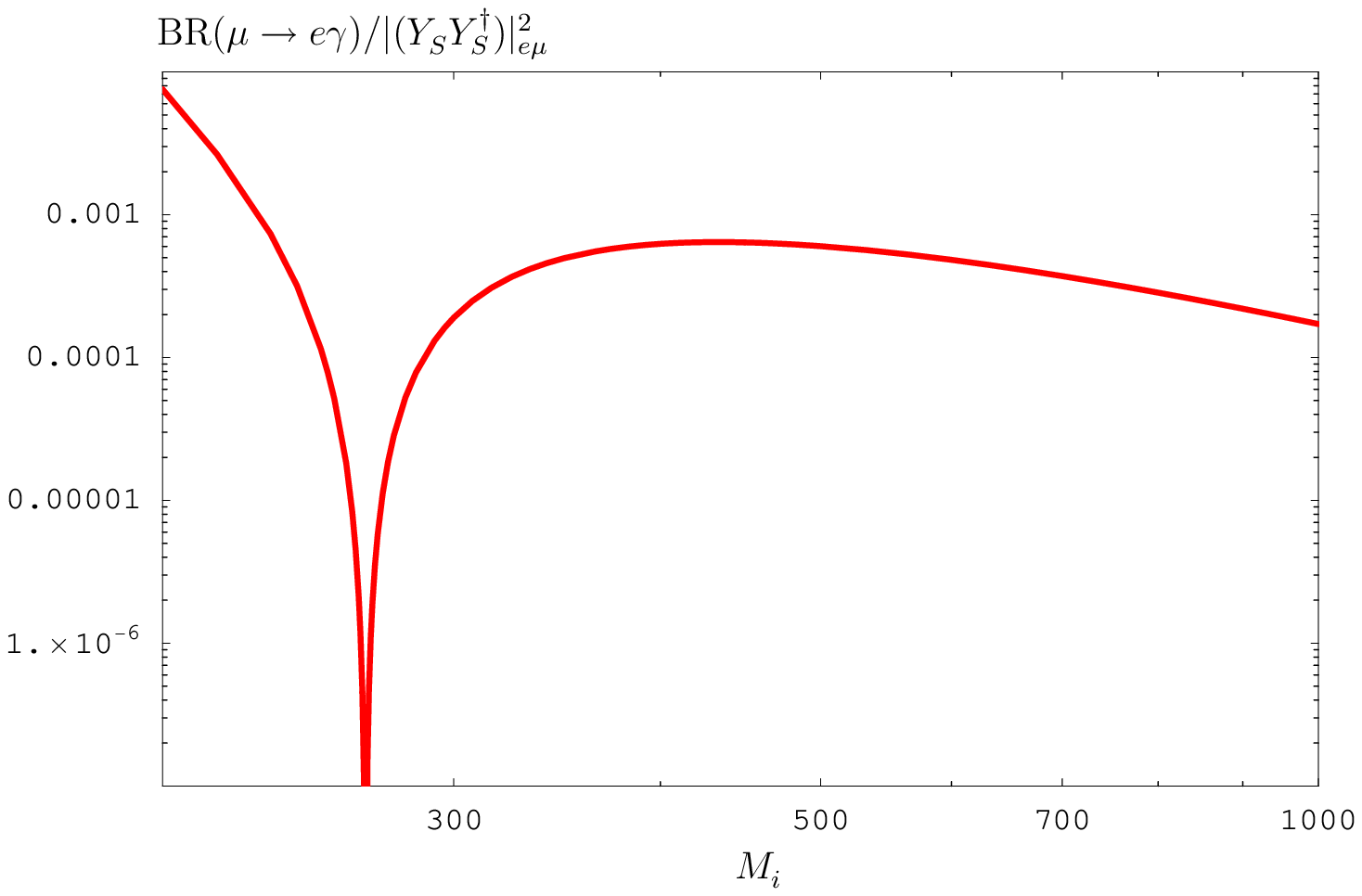,height=5.5cm,width=7.5cm,angle=0}}
\caption{Branching ratios for (a) $\mu \rightarrow 3 e$ and (b) $\mu
\rightarrow e \gamma$.}
\end{figure}

In Fig. 2 (a), we plot ${\rm BR} (\mu \rightarrow 3 e)$ as function
of $|UU^\dagger|_{e\mu}^2$. The horizon line stands for current
experimental constraints. Our result shows that, to meet the
experimental data, $|UU^\dagger|_{e\mu}^2$ must lie below $ 8.5
\times 10^{-6}$. Assuming that there is only one generation vector
like fermion, we plot, in Fig. 2 (b), ${\rm BR} (\mu \rightarrow e
\gamma)/ |Y_S^{} Y_S^\dagger|^2_{e \mu}$ as function of $M_S^{}$ by
setting $M_h^{}=120 ~{\rm GeV}$. We find that, to get big Yukawa
coupling $Y_S^{}$, $M_S^{}$ must lie around $270~ {\rm GeV}$ or be
heavier than several ${\rm TeV}$.

\subsection{Electromagnetic properties of heavy Majorana neutrinos}

The electromagnetic properties of Majorana neutrinos show up, in a
quantum field theory, as its interaction with the photon, and is
described by the following effective interaction vertex: ${\cal
L}_{\rm eff}^{} = \bar \psi \Gamma_\mu^{} \psi A^\mu$.  The most
general matrix element of ${\cal L}_{\rm eff}^{}$ between two
one-particle states, i.e., $\langle p', s'| J_\mu(0) | p, s \rangle
= \bar u_{s'} (p') \Gamma_\mu^{} u_{s}(p)$, which is consistent with
the Lorentz invariance, can be written as
\begin{eqnarray}
\bar u (p_2^{}, s_2^{}) \Gamma^\mu u(p_1^{}, s_1^{}) = \bar u
(p_2^{}, s_2^{}) \left[ {\cal E}^{}(q^2) \gamma^\mu - {\cal M}^{}
(q^2) i \sigma^{\mu\nu} q^\nu + {\cal H}^{}(q^2)
q^\mu\right] u(p_1^{}, s_1^{})&& \nonumber \\
+ \bar u (p_2^{}, s_2^{}) \left[ {\cal G}^{}(q^2) \gamma^\mu
\gamma^5- {\cal T}^{} (q^2) i \sigma^{\mu\nu} q^\nu \gamma^5 + {\cal
S}^{}(q^2) q^\mu \gamma^5 \right] u(p_1^{}, s_1^{}) && \; ,
\end{eqnarray}
where $q= p_2^{} -p_1^{}$. $ 2 M {\cal M} (0) $ and $ 2 M {\cal T}
(0)$ correspond to the magnetic moment and electro dipole moment of
heavy neutrinos, respectively.
\begin{figure}[t]
\subfigure[]{\epsfig{file=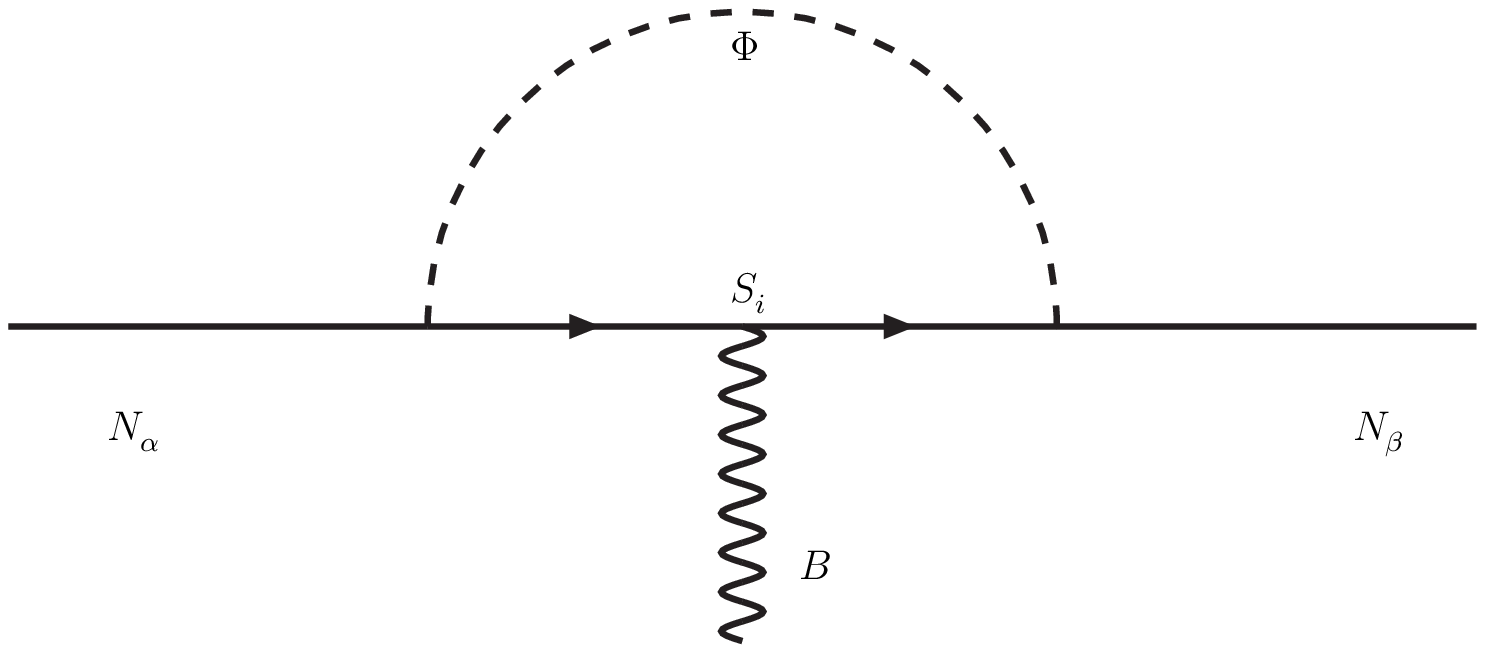,height=3cm,width=6cm,angle=0}}
\subfigure[]{\epsfig{file=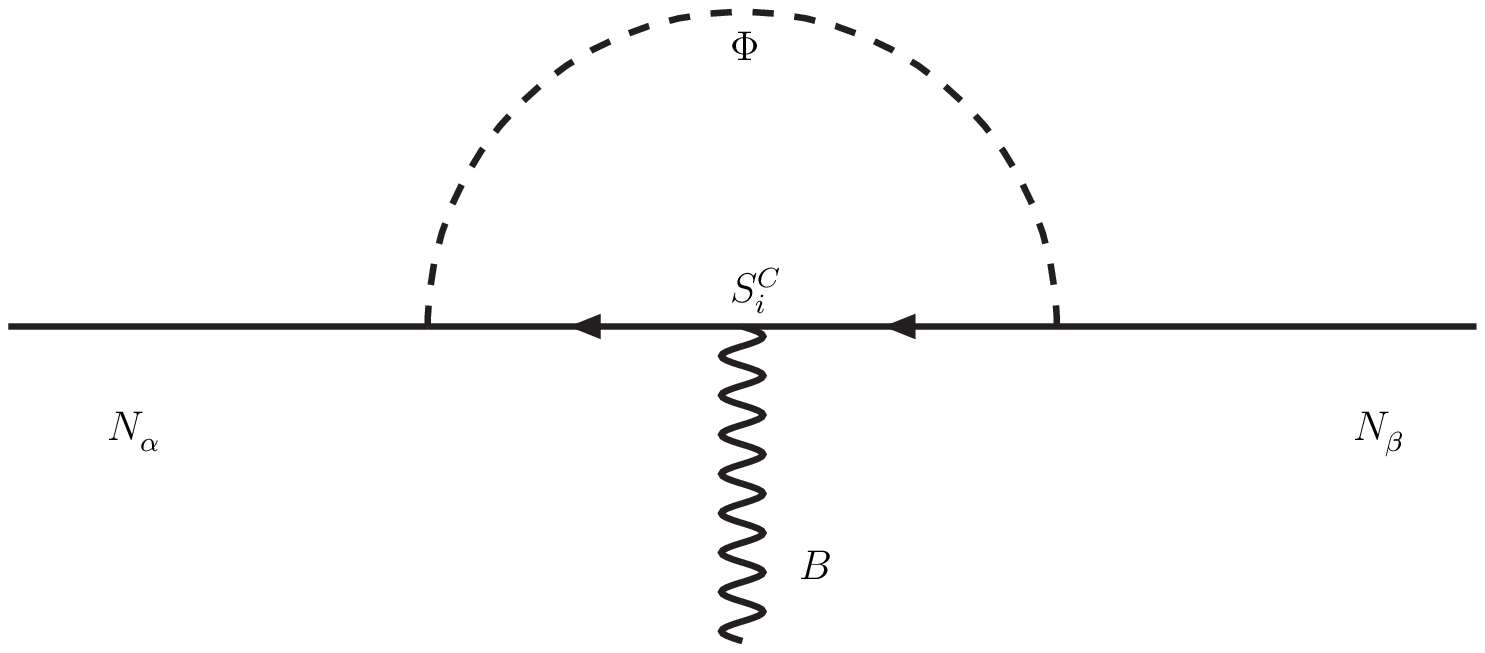,height=3cm,width=6cm,angle=0}}
\vspace{0.6cm}
\subfigure[]{\epsfig{file=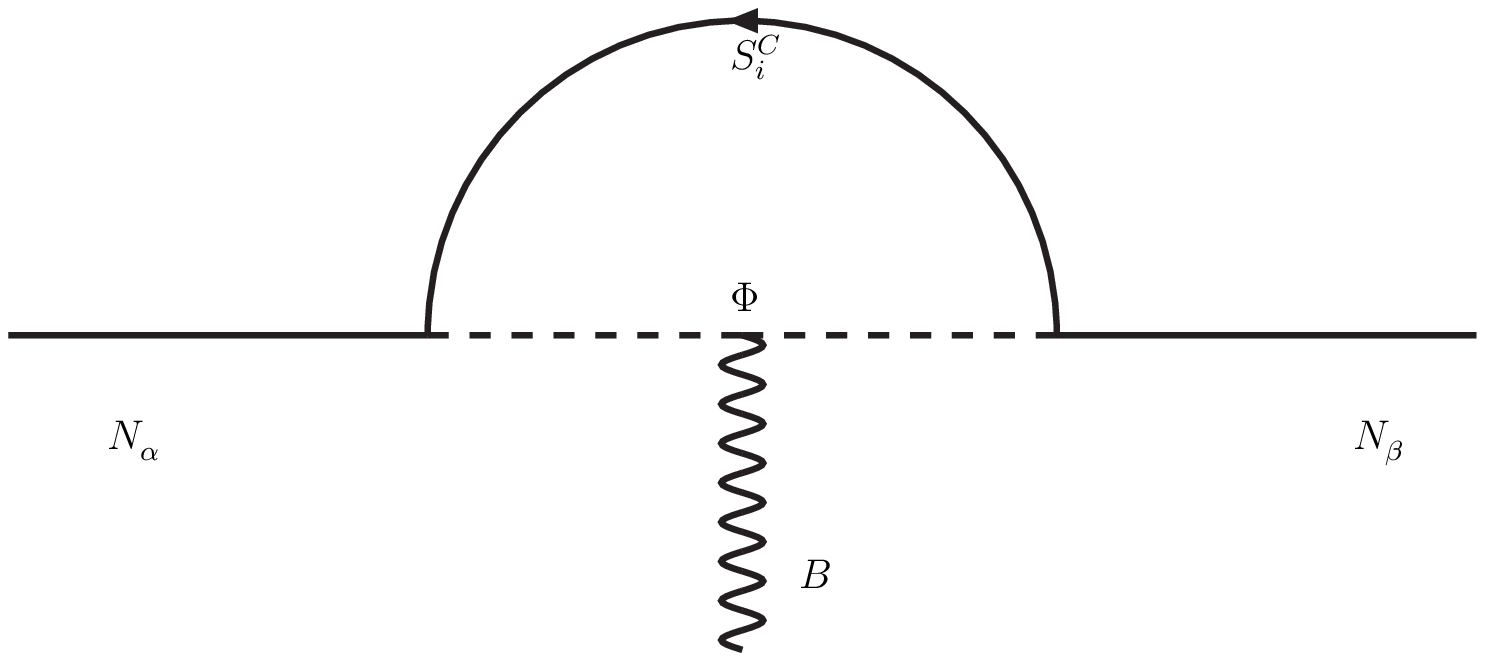,height=3cm,width=6cm,angle=0}}
\subfigure[]{\epsfig{file=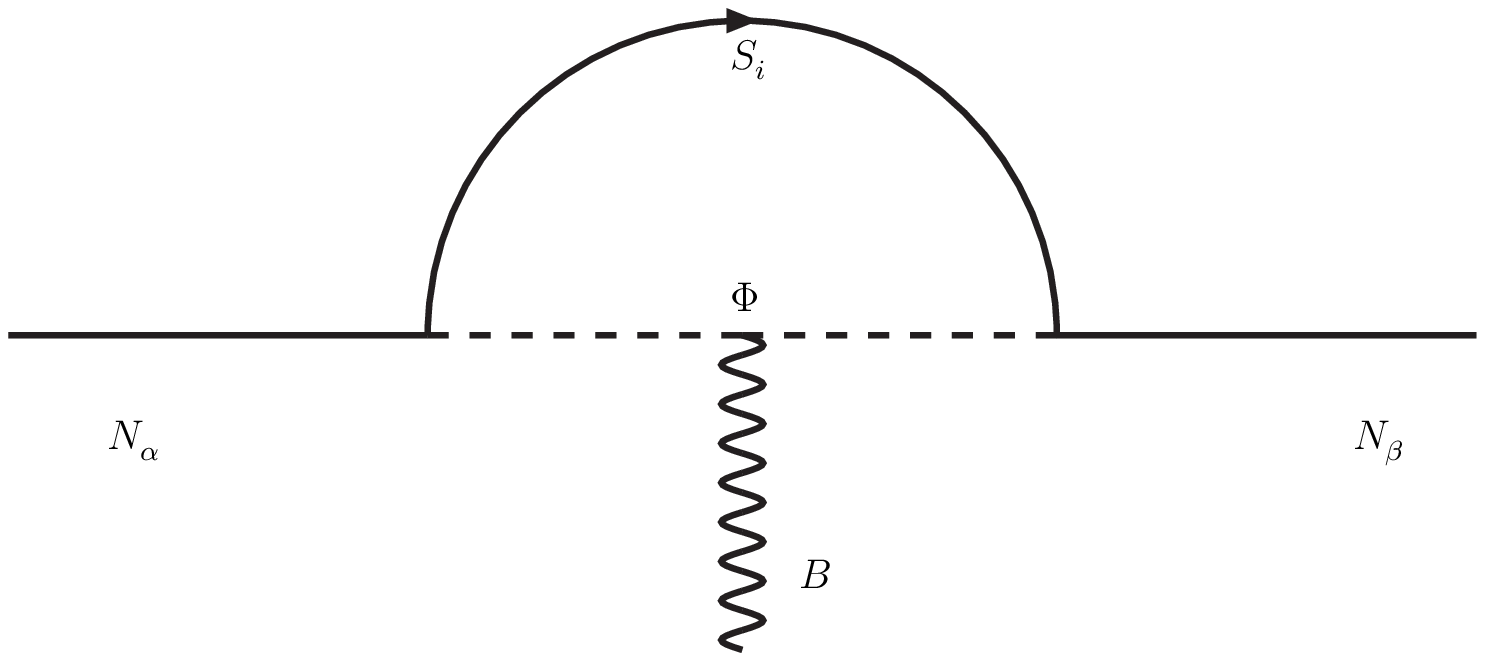,height=3cm,width=6cm,angle=0}}
\caption{Feynman diagrams contributing to heavy Majorana neutrino
transition magnetic moment.}
\end{figure}

Due to the Majorana nature, the magnetic moment of heavy Majorana
neutrinos is zero. There is only transition magnetic moment for
them. In the model considered, we have four diagrams contributing to
the transition magnetic moment, which are depicted in Fig. 3. The
Yukawa interactions of heavy Majorana neutrinos with $\Phi$ and $S$
can be rewritten in the following way
\begin{eqnarray}
{1 \over 2}\left[\overline{N^C_\alpha} \Phi^- (Y^T)_{\alpha i}^{}
P_R^{} S_i^C + \overline{ N_\alpha^{}} \Phi^+ (Y^\dagger)_{\alpha
i}^{} P_L^{} S_i^{}\right] + {1 \over 2 } \left[ \overline {S_i^{}}
Y_{i\alpha}^{} \Phi^- P_R^{} N_\alpha^{}  + \overline{S_i^C}
(Y^*)_{i \alpha}^{}  \Phi^+
 P_L^{} N_\alpha^C\right] \; ,
\end{eqnarray}
through which we can derive relevant feynman rules.  Assuming that
heavy Majorana neutrinos are nearly degenerate, i.e., $M_\alpha^{}
\approx M_\beta^{} \approx M$, we derive the transition magnetic
moment for heavy Majorana neutrinos
\begin{eqnarray}
{ a }_{\alpha \beta}^{N}= { M^2 \over 64 \pi^2}\left
[(Y_N^\dagger)_{\beta i}^{} (Y_N^{})_{i \alpha}^{} - (Y_N^T)_{\beta
i}^{} (Y_N^{})_{i \alpha}^{*} \right] \left[ {\cal I} (M_\Phi^2,
M^2, M_i^2) -{\cal  I} (M_i^2, M^2, M_\Phi^2)  \right] \; ,
\end{eqnarray}
with
\begin{eqnarray}
{\cal I}(A, B, C) = \int d x { x (1-x)^2 \over (1-x) A + x(x-1) B +
x C} \; , \nonumber
\end{eqnarray}
where $M_i^{}$ and $M_\Phi$ are the mass eigenvalues of heavy
vector-like fermion $S$ and scalar $\Phi$, respectively.

Now, we turn to some numerical analysis. As shown in Eq. (2),
$Y_N^{} \overline{S_L^{} } \Phi N_R^{}$ is totally the interaction
of new fields  beyond the SM, so that there is no experimental
constraint on $Y_N^{}$ except ${\cal O }(Y_N^{}) < \sqrt{4\pi} $ (to
satisfy the perturbation theory). We plot,  in Fig. 4, $|a_{\alpha
\beta}^N / [(Y_N^\dagger)_{\beta i}^{} (Y_N^{})_{i \alpha}^{} -
(Y_N^T)_{\beta i}^{} (Y_N^{})_{i \alpha}^{*} ] |$ as function of
$M_N^{}$. Assuming ${\cal O} ([(Y_N^\dagger)_{\beta i}^{}
(Y_N^{})_{i \alpha}^{} - (Y_N^T)_{\beta i}^{} (Y_N^{})_{i
\alpha}^{*} ])\sim 1 $, We can find that the transition magnetic
moment of heavy Majorana neutrinos can be of ${\cal O} (10^{-2})$
for special parameter settings. Our result in Eq. (23) is different
from that in Ref. \cite{wudka2} for not considering the Yukawa
coupling $\overline{N_R^{C}}\Phi S_R^{}$, which is forbidden by the
$Z_2^{}$ symmetry in our model. Given the large electromagnetic form
factors, heavy Majorana neutrinos can be produced at the LHC through
the electromagnetic interaction.

\begin{figure}[t]
{\epsfig{file=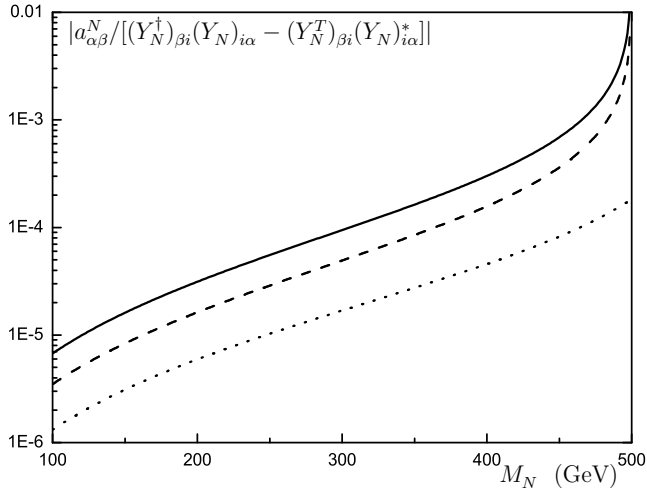,height=6.5cm,width=8.5cm,angle=0}}
\caption{Numerical illustration for the transition magnetic moment
of heavy Majorana neutrinos as function of $M_N^{}$. We assume there
is only one generation vector-like fermion $S$ and choose
$(M_\Phi^{}, ~M_i^{} )= (350, ~ 150) ~({\rm GeV})~$for solid line,
$(300, ~ 200) ~({\rm GeV})$ for dashed line and $(300, ~ 250) ~({\rm
GeV})$ for dotted line. }
\end{figure}

\subsection{Collider signatures}

We switch to comment on the collider signatures of our model. All of
the new particles introduced in the model lie around several hundred
GeV. Singly charged scalar and vector like fermions can be produced
through the electromagnetic interaction at the LHC. The most
promising production channel may be $ pp\rightarrow S^+ S^-
\rightarrow \ell^+ \ell^-  jjjj$ for heavy charged fermions and
$pp\rightarrow \Phi^+ \Phi^- \rightarrow jjjj$ for heavy charged
scalar. The production cross sections for these charged particles at
the LHC (with $\sqrt{s} = 14 {\rm TeV}$) are about several $fb$ when
heavy particle masses lie around 300 GeV \cite{wudka1, wudka2}. The
large transition magnetic moment can help to produce the heavy
Majorana neutrinos at the LHC. Its signatures are similar to that in
Type-III seesaw model \cite{tripletp1, tripletp2, tripletp3}. The
only distinguish is that, heavy neutrinos can not be produced
through weak interactions and must be produced in pair in our model.

\section{concluding remarks}

In this paper, we have proposed a novel TeV-scale seesaw mechanism.
One salient feature of our model is that Dirac neutrino mass matrix
raises only at the loop level. As a result, the heavy Majorana
neutrinos can be several hundred GeV. Another salient feature is
that heavy Majorana neutrinos can get large electromagnetic form
factors, through which they can be produced and detected at the LHC.
We have derived light Majorana neutrino mass formula and calculated
constraints on parameter space from LFV processes. At last we have
discussed the signatures of heavy fermions (vector-like fermions and
heavy Majorana neutrinos) and scalar at the LHC.

\begin{acknowledgments}
The author thanks to Tong Li, Yi Liao and Shu Luo for useful
discussion. This work was supported in part by the National Natural
Science Foundation of China.
\end{acknowledgments}

\appendix

\section{An alternative setting to the model}

Beside the model presented in section II, we can extending the SM
with different particle contents, which may lead to the same TeV
seesaw mechanism. For example, we can substitute $S_L^{}, S_R^{}$,
$\Phi$, with vector like fermion triplets $\Sigma_L^{}, \Sigma_R^{}$
and scalar triplet $\Delta$. In this case the lagrangian can be
written as
\begin{eqnarray}
{\cal L} = {\cal L}_{\rm SM}^{} -  \overline{\ell_{ L}^{} }
Y_\Psi^{} \tilde{H} \Sigma_R^{} - \overline{\Sigma_L^{}}
M_\Sigma^{}\Sigma_R^{}   - Y_N^{} {\rm Tr} [\Psi_L^{} \Delta] N_R^{}
+ V(H, \Delta) \; .
\end{eqnarray}
Here the weak hypercharge of the $\Psi$ and $\Sigma$ are zero.
$\Psi_L^{}$, $\Phi$ and $N_R^{}$ are odd , while the other fields
are even under $Z_2^{}$ transformation.

The Higgs potential can be written as
\begin{eqnarray}
V&=& {1 \over 2 } m_H^2 H^\dagger H + {1 \over 2} m_\Delta^{} {\rm
Tr}[ \Delta^\dagger \Delta] + {1 \over 4} \lambda (H^\dagger H)^2 +
\lambda_X^{} H^\dagger \Sigma i \sigma_2^{} H + \cdots \; ,
\end{eqnarray}
where dots denote Higgs potential terms we don't concern.


\begin{thebibliography}{99}

\bibitem{sno}

SNO Collaboration, Q. P. Ahamed {\it et al}., Phys. Rev. Lett. {\bf
89}, 011301 (2002).

\bibitem{abcc}

For a review, see: C. K. Jung {\it et al}., Ann. Rev. Nucl. Part.
Sci. {\bf 51}, 451 (2001).

\bibitem{kamland}

KamLAND Collaboration, K. Eguchi {\it et al}., Phys. Rev. Lett. {\bf
90}, 021802 (2003).

\bibitem{k2k}

K2K Collaboration, M. H. Ahn {\it et al}., Phys. Rev. Lett. {\bf
90}, 041801 (2003).

\bibitem{weinberg}

S. Weinberg, Phys. Rev. Lett. {\bf 43}, 1566 (1979).

\bibitem{seesawI}
P.~Minkowski,
  Phys.\ Lett.\ B {\bf 67}, 421 (1977);
  T.~Yanagida, in {\it Workshop on Unified Theories}, KEK report 79-18 p.95 (1979);
  M.~Gell-Mann, P.~Ramond, R.~Slansky,
  in {\it Supergravity} (North Holland, Amsterdam, 1979)
  eds. P.~van~Nieuwenhuizen, D.~Freedman, p.315;
  S.~L.~Glashow, in {\it 1979 Cargese Summer Institute on Quarks and Leptons} (Plenum Press,
  New York, 1980) eds. M.~Levy, J.-L.~Basdevant, D.~Speiser, J.~Weyers, R.~Gastmans and M.~Jacobs,
  p.687;
  R.~Barbieri, D.~V.~Nanopoulos, G.~Morchio and F.~Strocchi,
  Phys.\ Lett.\ B {\bf 90}, 91 (1980);
  R.~N.~Mohapatra and G.~Senjanovic,
  Phys.\ Rev.\ Lett.\  {\bf 44}, 912 (1980);
  G.~Lazarides, Q.~Shafi and C.~Wetterich,
  Nucl.\ Phys.\  B {\bf 181}, 287 (1981).

\bibitem{seesawII}
W.~Konetschny and W.~Kummer,
  Phys.\ Lett.\  B {\bf 70}, 433 (1977);
%
 T.~P.~Cheng and L.~F.~Li,
  Phys.\ Rev.\  D {\bf 22}, 2860 (1980);
%
 G.~Lazarides, Q.~Shafi and C.~Wetterich,
 Nucl.\ Phys.\  B {\bf 181}, 287 (1981);
%
 J.~Schechter and J.~W.~F.~Valle,
  Phys.\ Rev.\  D {\bf 22}, 2227 (1980);
%
 R.~N.~Mohapatra and G.~Senjanovic,
  Phys.\ Rev.\  D {\bf 23}, 165 (1981).


\bibitem{seesawIII}
R.~Foot, H.~Lew, X.~G.~He and G.~C.~Joshi,
  Z.\ Phys.\  C {\bf 44}, 441 (1989).


\bibitem{ma}

E. Ma and U. Sarkar, Phys. Rev. Lett. {\bf 80}, 5716 (1998).

\bibitem{Pilaftsis}

M. Y. Keung and G. Senjanovic, Phys. Rev. Lett. {\bf 50}, 1427
(1983); A. Pilaftsis, Z. Phys. C {\bf 55}, 275 (1992); B. Bajc, M.
Nemevsek and G. Senjanovic, Phys. Rev. D {\bf 76}, 055011 (2007).

\bibitem{early} J. Bernabeu, A. Santamaria, J. Vidal, A. Mendez, and
J.W.F. Valle, Phys. Lett. B {\bf 187}, 303 (1987); W. Buchmuller and
D. Wyler, Phys. Lett. B {\bf 249}, 458 (1990); W. Buchmuller and C.
Greub, Nucl. Phys. B {\bf 363}, 345 (1991); A. Datta and A.
Pilaftsis, Phys. Lett. B {\bf 278}, 162 (1992); G. Ingelman and J.
Rathsman, Z. Phys. C {\bf 60}, 243 (1993); C.A. Heusch and P.
Minkowski, Nucl. Phys. B {\bf 416}, 3 (1994); D. Tommasini, G.
Barenboim, J. Bernabeu, and C. Jarlskog, Nucl. Phys. B {\bf 444},
451 (1995).

\bibitem{smirnov}
J. Gluza, Acta Phys. Polon. B {\bf 33}, 1735 (2002); J. Kersten and
A.Yu Smirnov, Phys. Rev. D {\bf 76}, 073005 (2007); X. G. He, S. Oh,
J. Tandean and C. C. Wen, Phys. Rev. D {\bf 80}, 073012 (2009).



\bibitem{Han} T. Han and B. Zhang, Phys. Rev. Lett. {\bf 97}, 171804
(2006).

\bibitem{spanish}
F.del Aguila, J.A. Aguilar-Saavedra, A.M. de la Ossa, and M. Meloni,
Phys. Lett. B {\bf 613}, 170 (2005); F.del Aguila and J.A.
Aguilar-Saavedra, JHEP {\bf 0505}, 026 (2005); F.del Aguila, J. A.
Aguilar-Saavedra, and R. Pittau, JHEP {\bf 0710}, 047 (2007); N.
Haba, S. Matsumoto and K. Yoshioka, Phys. Lett. B {\bf 677}, 291
(2009).


\bibitem{tev type-II}
W. Chao, S. Luo, Z.Z. Xing, and S. Zhou, Phys. Rev. D {\rm 77},
016001 (2008); W. Chao, Z. Si, Z.Z. Xing, and S. Zhou, Phys. Lett. B
{\bf 666}, 451 (2008); Z. Z. Xing, Phys. Lett. B {\bf 679}, 255
(2009); W. Chao, Z. Si, Y. J. Zheng and S. Zhou, Phys. Lett. B {\bf
683}, 26 (2010).

\bibitem{zee}

A. Zee, Phys. Lett. B {\bf 93}, 389 (1980), Erratum-ibid.B {\bf 95},
461 (1980).

\bibitem{zerodis}

S. Antusch, C. Biggio, E. Fernandez-Martinez, M. B. Gavela and J.
Lopez-Pavon, JHEP {\bf 0610}, 084 (2006).

\bibitem{extest1}

E. Fernandez-Martinez, M. B. Gavela, J. Lepez-Pavon and O. Yasuda,
Phys. Lett. B {\bf 649}, 427 (2007).

\bibitem{zhizhong}

Z. Z. Xing, Phys. Lett. B {\bf 660}, 515 (2008).

\bibitem{extest2}

A. Abada, C. Biggio, F. Bonnet, B. Gavela and T. Hambye, JHEP, {\bf
0712}, 061 (2007).

\bibitem{wudka1}

A. Aparici, K. Kim, A. Santamaria and J. Wudka, Phys. Rev. D {\bf
80}, 013010 (2010).

\bibitem{wudka2}

A. Aparici, K. Kim, A. Santamaria and J. Wudka, arXiv: 0911.4103
[hep-ph].

\bibitem{tripletp1}

R. Franceschini, T. Hambye and A. Strumia, Phys. Rev. D {\bf 78},
033002 (2008).

\bibitem{tripletp2}

F. del Agulia and J. A. Aguilar-Saavedra, Nucl. Phys. B {\bf 813},
22 (2009).

\bibitem{tripletp3}

Tong Li and X. G. He, Phys. Rev. D {\bf 80}. 093003 (2009).



\end{thebibliography}
\end{document}